Природа дисперсии поляроных полос в спектре ARPES сверхпроводящих купратов: дополнение принципа Франка-Кондона

А. Э. Мясникова, Д. В. Мосейкин

Физический факультет, Южный федеральный университет, 344090 г.Ростов-на-Дону,.
Зорге 5
e-mail: myasnikova67@yandex.ru

Спиновые корреляции полагаются в настоящее время единственным возможным источником дисперсии поляронной полосы в спектре ARPES купратов. Однако, такая же дисперсия и большая ширина полосы характерна и для O $2p_\pi$ полосы в спектре ARPES купратов [Phys. Rev. B 75, 075115 (2007)], хотя O $2p_\pi$ состояния не связаны с низкоэнергетическими спиновыми степенями свободы. Цель настоящего доклада – исследование дисперсии поляронной полосы в спектре ARPES в рамках модели полярона Пекара, где она рассчитывается аналитически [Phys. Rev. B 77, 165136 (2008)]. Мы уточняем метод расчета учетом импульса отдачи и получаем новое аналитическое выражение для расчета поляронной полосы в спектре ARPES. Полученная в соответствии с ним полоса хорошо согласуется с поляронной полосой в спектре ARPES купратов. Дисперсия рассчитанной полосы обусловлена импульсом отдачи. Показано, что по дисперсии поляронной полосы в экспериментально полученном спектре ARPES можно определить эффективную массу полярона. Обсуждается необходимость дополнения принципа Франка-Кондона указанием, что в момент быстрого (фото)перехода не изменяются средние значения координат гармоник фононного поля, но изменяются их средние импульсы за счет импульса отдачи.

On the nature of dispersion of the polaronic bands in ARPES spectra of cuprates: supplement of the Franck-Condon principle

A. E. Myasnikova, D. V. Moseykin

*Physics Department, Southern Federal University, 5 Zorge str., 344090 Rostov-on-Don, Russia*

e-mail: myasnikova67@yandex.ru

Spin correlations are considered now as the only possible source of dispersion of the polaronic band in ARPES spectra of cuprates. However, the same dispersion is observed experimentally for broad O $2p_\pi$ band in ARPES spectra of cuprates [Phys. Rev. B 75, 075115 (2007)] although the O $2p_\pi$ states do not experience any coupling to the low-energy spin degrees of freedom. The aim of the present report is exploring dispersion of the polaronic band in ARPES spectrum in the frames of Frohlich-Pekar polaron model where this band is calculated analytically [Phys. Rev. B 77, 165136 (2008)]. We improve the method by taking into account the momentum of the recoil and obtain new analytical expression for the photoelectron rate as function of binding energy and momentum. Calculated according to it polaronic band demonstrates dispersion and broadening being in good agreement with the experiment. The dispersion is caused by the momentum of the recoil and can be used to extract the polaron effective mass from the experimental ARPES spectra. We also supplement the Franck-Condon principle by indication that if the phonon field is in a coherent state then during rapid (photo) transition the average coordinates of the phonon field harmonics do not change, but their average momentums are changed due to the recoil momentum.

Спектры ARPES купратов демонстрируют характерные широкие полосы с максимумом около значений энергии связи -0.4 ÷ -0.5 эВ, обусловленные, как это уже общепризнано, сильным электрон-фононным взаимодействием [1]. Однако для объяснения заметной дисперсии, которую демонстрируют эти полосы [1], до сих пор был предложен только один вариант, основанный на использовании t-J модели, расширенной учетом электрон-фононного взаимодействия. В этой модели полярон является поляроном малого радиуса, а дисперсия поляронной полосы в спектре ARPES совпадает с дисперсией "голого" носителя (т.е. без учета электрон-фононного взаимодействия). Таким образом, источником дисперсии поляронной полосы в спектре ARPES в этой модели является взаимодействие носителя заряда с низкоэнергетическими спиновыми степенями свободы.

Главное расхождение такой модели с экспериментально наблюдаемыми свойствами допированных купратов – способность поляронов малого радиуса к когерентному (зонному) перемещению только при температуре около абсолютного нуля. При более высоких температурах движение полярона малого радиуса осуществляется за счет активированных температурой перескоков. Такой характер движения носителей заряда не согласуется с наблюдаемыми свойствами допированных купратов, где носители заряда демонстрируют зонную проводимость.

Не менее важным является и второе расхождение модели, использующей поляроны малого радиуса с экспериментом. В такой модели не возникает сосуществование автолокализованных и делокализованных носителей заряда, которое наблюдается экспериментально (например, в спектрах ARPES можно одновременно наблюдать значительный спектральный вес около поверхности Ферми и при энергиях связи -0.8 ÷ -1.0 эВ). Наконец, третьим аргументом в пользу поиска альтернативной причины дисперсии поляронной полосы является тот факт, что аналогичную дисперсию демонстрирует и O $2p_\pi$ полоса в спектре ARPES купратов [1], хотя O $2p_\pi$ состояния не имеют никакой связи с низкоэнергетическими спиновыми степенями свободы.

Альтернативной моделью, позволяющей учесть сильное электрон-фононное взаимодействие, является модель полярона большого радиуса, или полярона Пекара. Он в отличие от полярона малого радиуса простирается на несколько элементарных ячеек. Как следствие, его можно описывать в континуальном приближении, а его движение является когерентным. Еще одним преимуществом полярона большого радиуса является то, что его фотоэмиссионный спектр рассчитывается аналитически

[2], в отличие от случая t-J модели [3]. Однако, существующий расчет предсказывает бездисперсионную полосу [2]. Мы в настоящем докладе показываем, что это является следствием недостаточно осторожного использования адиабатического приближения и получаем формулу для расчета поляронной полосы в спектре ARPES, демонстрирующую дисперсию, отлично согласующуюся с экспериментально наблюдаемой.

Действительно, расчет полосы в спектре ARPES, обусловленной фотодиссоциацией полярона, мы выполняем с помощью Золотого правила Ферми. Вектор начального состояния имеет вид

$$|i\rangle = \psi(\mathbf{r})\prod_{\mathbf{q}}|d_{\mathbf{q}}\rangle \ , \ \psi(\mathbf{r}) = \sqrt{7\pi\beta^{-3}}A(1+\beta r)e^{-\beta r}, \qquad (1)$$

где $\psi(\mathbf{r})$ - волновая функция носителя заряда в поляроне, а $\prod_{\mathbf{q}}|d_{\mathbf{q}}\rangle$ описывает состояние фононного поля в представлении когерентных состояний [4,5]. В этом представлении **q**-я гармоника характеризуется ненулевым средним значением операторов рождения/уничтожения фононов: $<b_q> = d_q, <b_q^+> = d_q^*$. Параметры $|d_q|, \varphi_q$ ($d_q = |d_q|e^{i\varphi_q}$) можно назвать параметрами деформации фононного вакуума. Их значения в покоящемся поляроне, как и значение параметра электронной волновой функции, определяются вариационным методом [4]:

$$|d_{\mathbf{q}}| = \frac{e}{q}\sqrt{\frac{2\pi}{V\varepsilon^*\hbar\omega_{\mathbf{q}}}}\eta_{\mathbf{q}}, \varphi_{\mathbf{q}} = -\mathbf{q}\mathbf{R}^{'}, \qquad (2)$$

где $\eta_q$ – фурье-образ квадрата электронной волновой функции, V – объем кристалла, $\varepsilon^*$ is – эффективная диэлектрическая прониаемость, характеризующа поляризуемость кристаллической решетки: $(\varepsilon^*)^{-1} = \varepsilon_\infty^{-1} - \varepsilon_0^{-1}$, $\varepsilon_0, \varepsilon_\infty$ - статическая и высокочастотная диэлектрические проницаемости, $\omega_\mathbf{q}$ - частота **q**-й гармоники фононного поля, **R'** - радиус-вектор центра полярона. Конечное состояние электрона $|f\rangle = \sqrt{V}\exp(i\mathbf{k}\mathbf{r})\prod_{\mathbf{q}}|\{v_\mathbf{q}\}\rangle$ описывается плоской волной, а конечное состояние фононного поля характеризуется определенным значением $\upsilon_\mathbf{q}$ числа фононов в каждой гармонике, которые возникли в результате распада когерентного состояния фононного поля в поляроне после того, как носитель заряда покинул сформированную им поляризационную потенциальную яму.

Очевидно, что электронная часть матричного элемента перехода, представляющего собой произведение электронной и фононной частей [5,2]:

$$\langle f|\hat{H}_{int}|i\rangle = \int d\mathbf{r} L^{-3/2}\exp(-i\mathbf{kr})\hat{H}_{int}\psi(\mathbf{r})\prod_{\mathbf{q}}\langle v_{\mathbf{q}}|d_{\mathbf{q}}\rangle \qquad (3)$$

выражает закон сохранения импульса носителя заряда при фотопереходе. Однако, поскольку полярон и фотон образуют замкнутую систему, сохраняться при фотодиссоциации должен суммарный импульс носителя заряда и его "фононной шубы", или деформации фононного вакуума, (в пренебрежении малым импульсом фотона). Таким образом, при применении Золотого правила Ферми к системам, рассматриваемым в адиабатическом приближении, нужно предпринять дополнительные меры для соблюдения сохранения полного импульса замкнутой системы. Для этого необходимо учесть импульс отдачи, уносимый "фононной шубой" полярона после того, как носитель ее покинул.

Если электрон покинул кристалл, имея проекцию импульса на плоскость поверхности кристалла $\hbar k_{||}'$, проекция импульса отдачи, уносимого деформацией фононного вакуума

$$p_{pvd} = -\hbar(k_{||}' - nK) \equiv -\hbar k_{||}, \qquad (4)$$

где K – вектор обратной решетки. В результате деформированный фононный вакуум приобретает кинетическую энергию

$$E_{kin} = p_{pvd}^{2}/2m_{pvd}^{*}, \qquad (5)$$

где $m_{pvd}^{*}$ - эффективная масса деформации фононного вакуума. Она легко может быть найдена по эффективной массе полярона $m_{pol}^{*}$ и носителя заряда $m^{*}$:

$$m_{pvd}^{*} = m_{pol}^{*} - m^{*}. \qquad (6)$$

Связанная с появлением ненулевого импульса кинетическая энергия фононного поля, как и его потенциальная энергия (равная ее значению в основном состоянии, полученному в результате минимизации), дает вклад в среднее значение энергии фононного поля, и следовательно, в среднее значение

$$\overline{v} = \sum_{\mathbf{q}}|d_{\mathbf{q}}|^{2} = (\overline{E}_{kin} + \overline{E}_{pot})/\hbar\omega \qquad (7)$$

оператора числа квантов, которое определяет и вид распределения Пуассона для вероятностей излучения каждого возможного значения числа фононов при фотодиссоциации полярона [5,2]:

$$P_{\nu} = \frac{\overline{v}^{\nu-1}}{(\nu-1)!}e^{-\overline{v}}, \qquad (8)$$

где $P_\nu = \sum_{\{\nu_\mathbf{q}\}} \prod_\mathbf{q} \left|\langle \nu_\mathbf{q} | d_\mathbf{q} \rangle\right|^2$ - квадрат матричного элемента перехода фононной подсистемы, просуммированный по всем событиям, в которых излучается одинаковое число фононов [5,2] (в приближении бездисперсионных фононов эти события соответствует одинаковой энергии излучения).

Вероятность фотодиссоциации полярона, выраженная как функция параллельной плоскости кристалла проекции волнового вектора и энергии электрона вне среды с излучением $\upsilon$ фононов имеет вид [2]

$$dW_{\nu,\mathbf{k}} = \frac{256 e^2}{7\pi \hbar m^* c^2 \beta^3} \frac{\left(k_{||} \cos\varphi \cos\psi + \sqrt{2m^*(\varepsilon' + \Phi)/\hbar^2 - k_{||}^2} \sin\psi\right)^2 A^2}{(1 + \beta^{-2}(k'^2 m^*/m_e + 2m^*\Phi))^{-6}} \cdot \frac{\bar{\nu}^{\nu-1}}{(\nu-1)!} e^{-\bar{\nu}} \mathrm{K} d\Omega'$$

$$\mathrm{K} d\Omega' = d\Omega' \frac{\varepsilon' \sqrt{2m_e \varepsilon'/\hbar^2}}{\varepsilon' + \Phi} \frac{\sqrt{2m_e \varepsilon'/\hbar^2 - k_{||}^2}}{\sqrt{2m^*(\varepsilon' + \Phi)/\hbar^2 - k_{||}^2}} \quad , \tag{9}$$

где $\bar{\nu}$ теперь определяется выражением (7) с учетом выражений (4-6). Поляронная полоса в спектре ARPES получается как огибающая парциальных вкладов (9) для каждого значения $\upsilon$ [2].

Таким образом, мы получили новое аналитическое выражение для расчета полосы в спектре ARPES, обусловленной фотодиссоциацией полярона Пекара при учете импульса отдачи. Пример рассчитанной в соответствии с этим выражением полосы приведен на Рис.1. Видно, что он находится в отличном согласии с экспериментом [1] как по дисперсии максимума, так и по изменению полуширины полосы.

Фотодиссоциация полярона обычно рассматривается как Франк-Кондоновский процесс, при котором состояние фононного поля не изменяется. В таком приближении закон сохранения импульса замкнутой системы, очевидно, нарушается. Поэтому при применении принципа Франка-Кондона к фотодиссоциации полярона его необходимо дополнить с учетом импульса отдачи следующим образом: в момент быстрого электронного фотоперехода не меняется среднее значение координаты гармоник фононного поля; значение среднего импульса гармоник фононного поля, находящегося в когерентном состоянии, изменяется в момент перехода в соответствии с законом сохранения импульса.

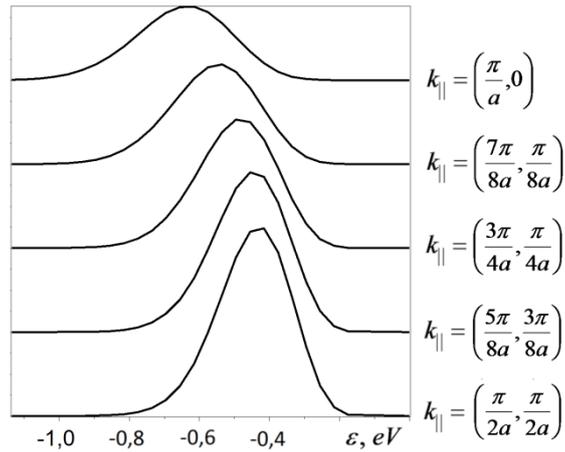

Рис.1.Рассчитанная при учете импульса отдачи дисперсия поляронной полосы в спектре ARPES.